\documentclass[10pt,sigconf,anonymous=false]{acmart}

\renewcommand\footnotetextcopyrightpermission[1]{} % removes conference notes
\usepackage{times}  
\usepackage{hyperref}
\usepackage[colorinlistoftodos]{todonotes}
\usepackage{subcaption}
\usepackage[printonlyused]{acronym}
\usepackage{tabularx}
\usepackage{balance}

\begin{document}

\acrodef{TAP}{Transferred Account Procedure}
\acrodef{SCCP}{Signalling and Connection Control Part}
\acrodef{DLT}{Distributed Ledger Technology}
\acrodef{xDR}{eXtended Detail Record}
\acrodef{TAC}{Type Allocation Code}
\acrodef{PoP}{Point of Presence}
\acrodef{IMEI}{International Mobile Equipment Identity}
\acrodef{EMSISDN}{Encrypted Mobile Station International Subscriber Directory Number}
\acrodef{MSISDN}{Mobile Station International Subscriber Directory Number}
\acrodef{MME}{Mobility Management Entity}
\acrodef{VoLTE}{Voice over \ac{LTE}}
\acrodef{MVNO}{Mobile Virtual Network Operators}
\acrodef{CDR}{Call Detail Record}
\acrodef{IMSI}{International Mobile Subscriber Identity}
\acrodef{RAT}{Radio Access Technology}
\acrodef{LPWA}{Low Power Wide Area}
\acrodef{VMNO}{Visited Mobile Network Operator}
\acrodef{HMNO}{Home Mobile Network Operator}
\acrodef{DCH}{Data Clearing House}
\acrodef{GSMA}{\ac{GSM} Association}
\acrodef{GSM}{Global System for Mobile communications}
\acrodef{TAP}{Transferred Account Procedure}
\acrodef{LTE-M}{\ac{LTE} Machine Type Communication}
\acrodef{MTC}{Machine Type Communications}
\acrodef{NB-IoT}{Narrow Band \ac{IoT}}
\acrodef{IOT}{Inter Operator Tariff}
\acrodef{IoT}{Internet of Things}
\acrodef{M2M}{Machine-to-Machine}
\acrodef{FNO}{Fixed Network Operator}
\acrodef{ISP}{Internet Service Providers} 
\acrodef{ASP}{Application Service Providers}
\acrodef{IPX}{IP Packet Exchange}
\acrodef{2G}{Second Generation}
\acrodef{3G}{Third Generation}
\acrodef{4G}{Fourth Generation}
\acrodef{5G}{Firth Generation}
\acrodef{ADB}{Android Debug Bridge}
\acrodef{ASCI}{Advertising Standards Council of India}
\acrodef{ASN}{Autonomous System Number}
\acrodef{CDN}{Content Delivery Network}
\acrodef{DL}{Downlink}
\acrodef{DNS}{Domain Name Service}
\acrodef{EaaS}{Experiment as a Service}
\acrodef{ECDF}{Empirical Cumulative Distribution Function}
\acrodef{HTTP}{Hyper Text Transfer Protocol}
\acrodef{ICMP}{Internet Control Message Protocol}
\acrodef{ISP}{Internet Service Provider}
\acrodef{IXP}{Internet Exchange Point}
\acrodef{LTE}{Long Term Evolution}
\acrodef{MBB}{Mobile Broadband}
\acrodef{MSC}{Message Sequence Chart}
\acrodef{E2E}{End-to-end}
\acrodef{QCI}{QoS Class Identifier}
\acrodef{NDT}{Network Diagnostic Tool}
\acrodef{QoE}{Quality of Experience}
\acrodef{QoS}{Quality of Service}
\acrodef{OS}{Operating System}
\acrodef{APN}{Access Point Name}
\acrodef{RMBT}{RTR Multithreaded Broadband Test}
\acrodef{RTT}{Round-Trip Time}
\acrodef{SIM}{Subscriber Identity Module}
\acrodef{SIMs}{Subscriber Identity Modules}
\acrodef{SLA}{Service-Level Agreement}
\acrodef{TCP}{Transmission Control Protocol}
\acrodef{UDP}{User Datagram Protocol}
\acrodef{UE}{User Equipment}
\acrodef{UL}{Uplink}
\acrodef{NRA}{National Regulatory Authority}
\acrodef{EC}{European Commission}
\acrodef{SGW}{Serving Gateway}
\acrodef{PGW}{Packet Data Network Gateway}
\acrodef{GTP}{GPRS Tunneling Protocol}
\acrodef{MNO}{Mobile Network Operator}
\acrodef{EU}{European Union}
\acrodef{HR}{home-routed roaming}
\acrodef{LBO}{local breakout}
\acrodef{IHBO}{IPX hub breakout}
\acrodef{VoIP}{Voice over IP}
\acrodef{IPG}{Inter-Packet Gap}
\acrodef{KS}{Kolmogorov-Smirnov}
\acrodef{FQDN}{Fully Qualified Domain Name}
\acrodef{MNC}{Mobile Network Code}
\acrodef{MCC}{Mobile Country Code}
\acrodef{MCCMNC}{\ac{MCC}-\ac{MNC}}
\acrodef{SIP}{Session Initiation Protocol}

% Turns comments ON or OFF
\newif\ifcomment
%\commenttrue
\commentfalse

% Comments (or NOT)
\ifcomment
    % Add per author NOTE with counter
    \newcounter{DPNumberOfComments}
    \stepcounter{DPNumberOfComments}
    \newcommand{\dpnote}[1]{\textcolor{green}{\small \bf [DP\#\arabic{DPNumberOfComments}\stepcounter{DPNumberOfComments}: #1]}}
    \newcounter{ALNumberOfComments}
    \stepcounter{ALNumberOfComments}
    \newcommand{\ael}[1]{\textcolor{purple}{\small \bf [ael\#\arabic{ALNumberOfComments}\stepcounter{ALNumberOfComments}: #1]}}
    \newcommand{\NOTE}[1]
    {
      {\footnotesize\it
        \begin{center}
          \begin{tabular}{|c|}
           \hline
            \parbox{0.85\columnwidth}{
              \medskip
              #1
              \medskip} \\
            \hline
          \end{tabular}
        \end{center}
        }
    }
\else
    \newcommand\dpnote[1]{}
    \newcommand\ael[1]{}
    \newcommand\NOTE[1]{}
\fi

%%%%%%%%%%%% THIS IS WHERE WE PUT IN THE TITLE AND AUTHORS %%%%%%%%%%%%

\title{DICE: Dynamic Interconnections for the Cellular Ecosystem}

\author{Andra Lutu}
\affiliation{
	\institution{Telefonica Research}
}
\email{andra.lutu@telefonica.com}

\author{Marcelo Bagnulo}
\affiliation{
	\institution{University Carlos III of Madrid}
}
\email{marcelo@it.uc3m.es}

\author{Diego Perino}
\affiliation{
	\institution{Telefonica Research}
}
\email{diego.perino@telefonica.com}

%\maketitle

%%%%%%%%%%%%%  ABSTRACT GOES HERE %%%%%%%%%%%%%%

\begin{abstract}
To enable roaming of users, the cellular ecosystem integrates many entities and procedures, including specific infrastructure to connect \acp{MNO},  business partnerships or the use of third-party \acp{DCH} for billing.
Many of these rely on specifications rooted in dated and arcane practices, involving long waiting periods for financial clearing, complex billing models, and disparate mechanisms for dealing with inter-MNO disputes. 
In this paper, we propose a novel solution -- DICE (Dynamic Interconnections for the Cellular Ecosystem) -- aimed at facilitating dynamic collaboration between MNOs, and sustain fluid interconnection models between the end-users and MNOs.
DICE uses distributed ledger technology (DLT) to enable MNOs to interact directly, and offer customizable services to their users through the use of crypto-currencies.
We leverage real-world data from a major operational MNO in Europe to support our claims, and to extract the requirements for the DICE system. 
We introduce the DICE protocol, and discuss real-world implementation considerations. 
\end{abstract}
\maketitle

%%%%%%%%%%%%% INTRODUCTION  %%%%%%%%%%%%%%

\section{Introduction}
\label{sec:introduction}
% !TeX root = paper.tex
% ================================================================

Roaming is one of the fundamental features of cellular networks. 
It enables clients of one \ac{MNO} to use the network of another \ac{MNO} when traveling outside their provider's area of coverage. 
To support roaming, partnering \acp{MNO} must have a commercial agreement in place, and a technical solution for exchanging signaling and voice/data traffic.
Despite the fast evolution of cellular technology over the past decades, this logic has remained largely unchanged. 
Even more, the associated platforms and systems are opaque, and little innovation has gone into this area. 
The advance of the inter-provider charging system is especially challenging, as it requires standardization, and subsequent joint evolution and deployment in the networks involved. 
This is inefficient, as the current inter-provider charging system for roaming services imposes penalties in terms of performance, operational costs and business revenues~\cite{gsma-interconnection}. %, which we describe next.
 
In this paper, we propose DICE (Dynamic Interconnections for the Cellular Ecosystem), a novel framework for dynamically establishing roaming agreements between MNOs, and enabling almost real-time financial clearing for roaming services through secure and private transfer of crypto-currencies.  
Most \acp{MNO} today choose to outsource to a \ac{DCH} for end-to-end roaming operations management, including data collection, clearing and cash settlement, although it is possible to do this on a peer-to-peer basis.
Even with \ac{DCH}, the charging settlement between MNOs remains a slow and tedious procedure (i.e., monthly or yearly for roaming partners with little traffic) and disputes are largely handled case-by-case. 
Based on real-world data from an MNO in Europe, this is a significant challenge for operators, which are grappling to remain relevant in an otherwise fast evolving landscape~\cite{UK_MNO_roaming}.
When analyzing the business interactions, \acp{MNO} have reported major revenue leakage, mainly because of incomplete customer records, accumulated roaming records, debt write-off and fraud. 
This issue is imperative, as support for ``things'' roaming internationally (e.g., connected cars, logistics and wearables) is critical for \ac{IoT} verticals.
Taking advantage of international mobile roaming, \ac{IoT} platforms can offer more stable connectivity/coverage services to IoT verticals. 
In this context, there is a stringent need for MNOs to develop sustainable strategies for next-generation interconnections to remain relevant.

The purpose of DICE is to allow MNOs to exchange value easily, without the need of a third party to act as a trusted intermediary, to verify the interaction between the roaming partners. 
With DICE, MNOs can avoid the need for using a DCHs and instead leverage the potential of \ac{DLT} and tokens to retrieve revenue from their roaming partners.

The current business logic around MNO interworking for roaming has further implications in the communication performance. 
The intrinsic lack of trust between the \ac{HMNO} and the \ac{VMNO}, and the unwillingness of the former to expose to a foreign operator charging information for their users makes \ac{HR} roaming the default roaming configuration.
Recent measurement studies confirmed that majority of \acp{MNO} deploys HR~\cite{mandalari2018experience}.
With HR, the roaming traffic is routed through the home country, regardless the roaming user's actual geo-location. 
%This implies that when the roaming user accesses services inside the visited country, 
%all packets travel twice between the visited and the home country~\cite{mandalari2018experience}.
This allows \acp{MNO} to control the charging function of their outbound roaming users, but is also translates into a non-negligible delay penalty and potential performance impairment, as previously uncovered via measurements in the wild~\cite{mandalari2018experience}. 
With existing regulation meant to tackle silent roamers and boost roaming revenues (e.g., Roam like at Home in Europe~\cite{ec-roaming-charges}), users also expect high quality, always-on services in a visited network; if these expectations are not met the risk of churn increases. 
%There is thus an urgency for MNOs to develop sustainable and dynamic roaming strategies. 
Facilitating \ac{LBO} roaming and simplifying the inter-MNO charging settlement would respond to this innovation need. 
However, all these require \acp{MNO} to trust their roaming partners to correctly charge roamers, without leaving the doubt that fraud (e.g., tampering with roaming records) might occur. 
The shift in the business logic that DICE brings through the use of crypto-currencies deliver this, and thus promotes the local breakout roaming configuration. 
We focus our design on data communications and LTE technology\footnote{For voice traffic, there is the need for a permanent identifier which depends on the \ac{HMNO}, so incoming calls will always be routed through the HMNO. We are not changing this logic, instead we focus on data communications (and outgoing calls) that only require the roamer to have an attachment point to the visited network}. 
%\dpnote{Should we say in the above paragraph private DLT? or in general mention we aim for a consortium/private DLT?}

With DICE, we make a two-fold contribution to disrupt the cellular ecosystem: 
(i) we enable MNOs to have a direct dynamic cooperation in a private and secure setup, establish roaming relationships, and perform data and financial clearing in almost real time, and 
(ii) we enable the end-user to have control over her mobile connection and connect to a network in a visited country as a native user, receiving optimal service performance.
The \ac{DLT} solution lowers uncertainty about the identity of the different entities involved in the roaming transaction (i.e, the roaming user, the home network and the visited network offering roaming services to the roaming user), and allows the exchange of value without trust between the entities. 
The purpose of DICE is to allow MNOs to engage in a roaming partnership and to exchange value easily, without the need of a third party to act as a trusted intermediary with the role to verify the interaction between the roaming partners. 
The notion of uncertainty we consider integrates three main components: identity of partner entities, visibility of transactions between any entities, and recourse in case something goes wrong. 
%\dpnote{make language a bit more crisp. Like "to provide the above requirements either there is a third party, or other solutions like databases etc. do not work as there is not a single entity that can be trusted".}
No individual organization in the group can be trusted with maintaining this archive of records, because falsified or deleted information would significantly benefit that party while damaging the others. 
Nonetheless, it is vital that all agree on the archive's contents, in order to prevent charging disputes.

\section{Background and Related Work}
\label{sec:background}
% !TeX root = paper.tex
% ================================================================

%\begin{figure}[t]
%\centering
%\includegraphics[width=0.5\textwidth]{fig/example.png}
%\vspace{-6mm}
%	\caption{Example of a user roaming across MNOs.}
%	\label{fig:example}
%	\vspace{-5mm}
%\end{figure}

\subsection{Roaming Background}
%\dpnote{Add in this section the private conversation bit again.}

%As a preliminary step,  HMNO and VMNO have to set up a roaming agreement to enable their customers to roam across them. 
%This agreement contains both the technical and commercial components required to enable the service, and GSMA provides guidelines for standard format for its members. 

%Roaming enables clients of one \ac{MNO} to use the network of another \ac{MNO} 
%when traveling outside their provider's area of coverage. 
To support roaming, the two partners must have a functioning technical solution 
and a commercial agreement in place (i.e., valid interworking). 
%MNOs negotiate agreements, implement and test 
%the roaming technical solution of choice and enable interworking. Interworking is 
%defined as the entirety of technical, commercial and operational means necessary to 
%join the service platforms of two \acp{MNO}, including all parties in the chain, 
%to enable a minimum of two end users to use a specific service offered by both 
%\acp{MNO} across their network boundaries. 
Once interworking is established, the Home Subscriber Servers (HSS) of the roaming partners communicate among them when a roaming customer (Alice) of an HMNO tries to authenticate to the VMNO. 

When Alice is outside her provider's area of coverage, she tries to connect to a local \ac{MNO} according to a list of preferred \acp{MNO} per geographical area provided by HMNO. 
Assuming VMNO is selected, roamer Alice follows the specified procedure. %described in Fig.~\ref{fig:example}. 
First, it contacts the HSS of VMNO with their international mobile subscriber identity number (IMSI). 
The HSS recognizes that Alice's IMSI does not belong to VMNO, but rather to HMNO and checks for a roaming agreement. 
If the agreement is in place, the HSS contacts its counterpart in HMNO, otherwise it refuses the access. 
This procedure leverages an international signaling infrastructure that is usually provided by a third party. 
%(e.g., \ac{SCCP} function of the \ac{IPX} in Figure~\ref{fig:example}). % named Signalling and Connection Control Part (SCCP) provider. 

Once authentication is completed there are three possible configurations for data transfer. 
The one that is currently deployed is \textit{\ac{HR}}~\cite{lte-epc-roaming-guidelines, lte-roaming-whitepaper}.
% All traffic to and from the mobile user is routed through the home network, for which a \ac{GTP} tunnel is set up between the \ac{SGW} of the visited network and the \ac{PGW} of the home network. With the IP end point in the home network, all services will be available in the same way as in the home network. This communication can be either direct or via an Internetwork Packet Exchange (IPX), this latter being the commonly deployed option.
Other two configurations that are not actually popular are \textit{local breakout (\ac{LBO})}~\cite{lte-epc-roaming-guidelines, lte-roaming-whitepaper} and  \textit{\ac{IHBO}}~\cite{US20140169286A1}. 
For the former, the visited network assigned the IP address of the roaming user; for the latter, the \ac{IPX} provides Alice's IP address.

With the most popular configuration (i.e., HR), HMNO logs Alice's activity at its \ac{PGW} while VMNO does the same at its \ac{SGW}. 
Both \acp{MNO} generate \ac{TAP} files containing Alice's activity for charging, which are collected by the \ac{DCH}. 
%These files could be exchanged in a peer-to-peer fashion but most MNOs choose to exchange these records via Data Clearing Houses (DCHs). 
The \ac{DCH} validates and verifies the \ac{TAP} files and in case of disagreement between information from \acp{MNO}, they trigger a dispute. 
The \ac{DCH} automatically handles disputes for minor differences, but in most cases disputes are handled case-by-case. 
The charging settlement procedure is not real time but rather happens yearly, monthly, or when the bill exceeds a certain threshold.

\subsection{Related Work}

Although prior work extensively investigated business aspects of the wired internet~\cite{lutu2012economic}, the interconection of networks~\cite{fanou2016pushing} and its evolution in time~\cite{dhamdhere2008ten}, the interactions between mobile networks and the global Internet remain largely unexplored. 
Some prior work has underlined structural characteristics of MNOs, internal organization~\cite{vallina2015beyond} and their impact around the world~\cite{rula2017cell}. 
However, we have little visibility on the mobile networks' interconnection dynamics and associated business relationships in the context of international roaming. 
This is an important gap to fill, especially since we know that the manner in which mobile networks interconnect can have important side-effects on the experience of the end-user~\cite{mandalari2018experience, UK_MNO_roaming}. 
In particular, an important effect comes from the relationships between mobile connectivity providers and content providers~\cite{michelinakis18infocom}.
The idea of leveraging distributed ledgers for increasing the dynamics of business agreements in the wired internet (i.e., at Internet Exchange Points) has recently been presented as part of the Dynam-IX framework~\cite{marcos2018dynam}. 
Similarly, DICE allows for roaming agreement updates in a dynamic manner. 
Additionally, DICE enables \ac{LBO} roaming and allows \acp{MNO} to avoid tedious procedures for data and financial clearing implemented in the current ecosystem through the creation of a roaming crypto-currency. 
Furthermore, the use of \ac{DLT} enables DICE to eliminate the need for roaming fraud detection, as roaming records become immutable once committed. 
Similar benefits of \ac{DLT} have also been exploited in charging schemes for third-party infrastructure usage in the cellular edge~\cite{li2019bridging}. Though used in a different application than the one we present in this paper, this validates the feasibility of DLT-enabled solutions to work in a production environment and tackle hard problems such as interaction in the cellular ecosystem without trust.

\section{DICE Toy Example}
\label{sec:toy_example}
% !TeX root = paper.tex
% ================================================================

% the idea in this section is to provide a basic description of the simplest DICE solution that we envision

In this section, we provide an overview of how DICE innovates the data and financial settlement procedures, and  removes the performance issues for using HR roaming. %In the following we first discuss DICE design considerations.
%Note that we consider LTE and data communications.
%; then we present the basic DICE design via an example of how a roaming customer of a given \ac{MNO} (hereinafter, the Home Mobile Network Operator (HMNO)) can use the DICE for international roaming and access the network of a foreign \ac{MNO} (hereinafter, the Visited Mobile Network Operator (VMNO)). 

%Before all, we provide a concise overview of the current ecosystem of international roaming from the point of an operational MNO and prepare the motivation for using DICE. 
%Based on this input, we then present a basic scenario on how DICE could help a roaming user and simplify the associated procedures that enable international mobility of cellular users. 
%We specifically highlight how DICE innovates the data and financial settlement procedures while also removing the performance issues discussed in Sec.~\ref{sec:introduction} for using HR roaming. 
%Finally, we introduce the high level requirements of a systems addressing current limitations, and provide an overview of how DICE tackle these requirements.

\subsection{DICE Design Considerations}
%As described in the previous section, the revenue collection procedures currently deployed in the cellular ecosystem require that both roaming partners generate service logs for their roaming customers.
%They then exchange this information with a trusted third party that further provides clearing and settlement services. 
%This current model of interaction forces operators to implement HR as roaming routing configuration to keep visibility of their outbound roamers. 
%This translates into a performance penalty for the roaming users. 

DICE aims to facilitate dynamic collaboration between MNOs by: 
(i) eliminating the need for an intermediary, such as a \ac{DCH} for clearing and settlement services and 
(ii) supporting the implementation of \ac{LBO} roaming by enabling the roaming user to attach to a visited network as a native user.

To achieve these goals DICE must ensure the following:

\noindent \emph{Near real time billing}: DICE should enable \acp{MNO} to bill a user and settle the roaming charges among them in almost real time. 
This would guarantee the user is able to consume the traffic she is allowed, avoiding incomplete and accumulated roaming records.

\noindent \emph{Automated dispute resolution}: DICE should avoid the generation of disputes among MNOs, and automatically solve any billing issue arising. 

\noindent \emph{Trust}: DICE should provide mechanisms to build trust among MNOs to guarantee optimal performance for the customer (e.g., enable \ac{LBO} roaming).
%Today this is provided by the CH and the solution should avoid such a centralized approach.

\noindent \emph{Confidentiality}: DICE must guarantee that any information considered private by the MNOs (e.g., billing information across roaming partners, or personal customer information) does not leak to unauthorized parties. 

DICE builds on \ac{DLT}, also known as blockchain, to enable \acp{MNO} to interact directly and offer custom services to their users (e.g., through the use of crypto-currencies). 
\acp{MNO} in the DICE consortium are free to engage in dynamic and versatile business relationships, without the need of third-party facilitators. 
The components of the DICE framework include: i) a protocol to automate the interaction of mobile operators (also with their subscribers); ii) a template to specify contracts easily and the legal framework to handle them; iii) a blockchain solution to store every transaction between the involved entities and enable the exchange of value.
This enables \acp{MNO} to interact directly without trust, benefit from the immutability of roaming records in the blockchain and ensure an optimal experience to end-users. 
%\dpnote{say here clearly how the DLT design and the above design solve the issues we listed, and recall why DLT is the only option.}

We provide next a toy example to demonstrate how Alice can use the DICE solution to roam internationally, and how DICE mechanisms provides near real time billing and avoid disputes across \acp{MNO}.  
We then discuss how to leverage existing \ac{DLT} to implement such mechanisms and guarantee trust and confidentiality in Sec.~\ref{sec:deployment}.

\begin{figure}[t]
\centering
\includegraphics[width=0.5\textwidth]{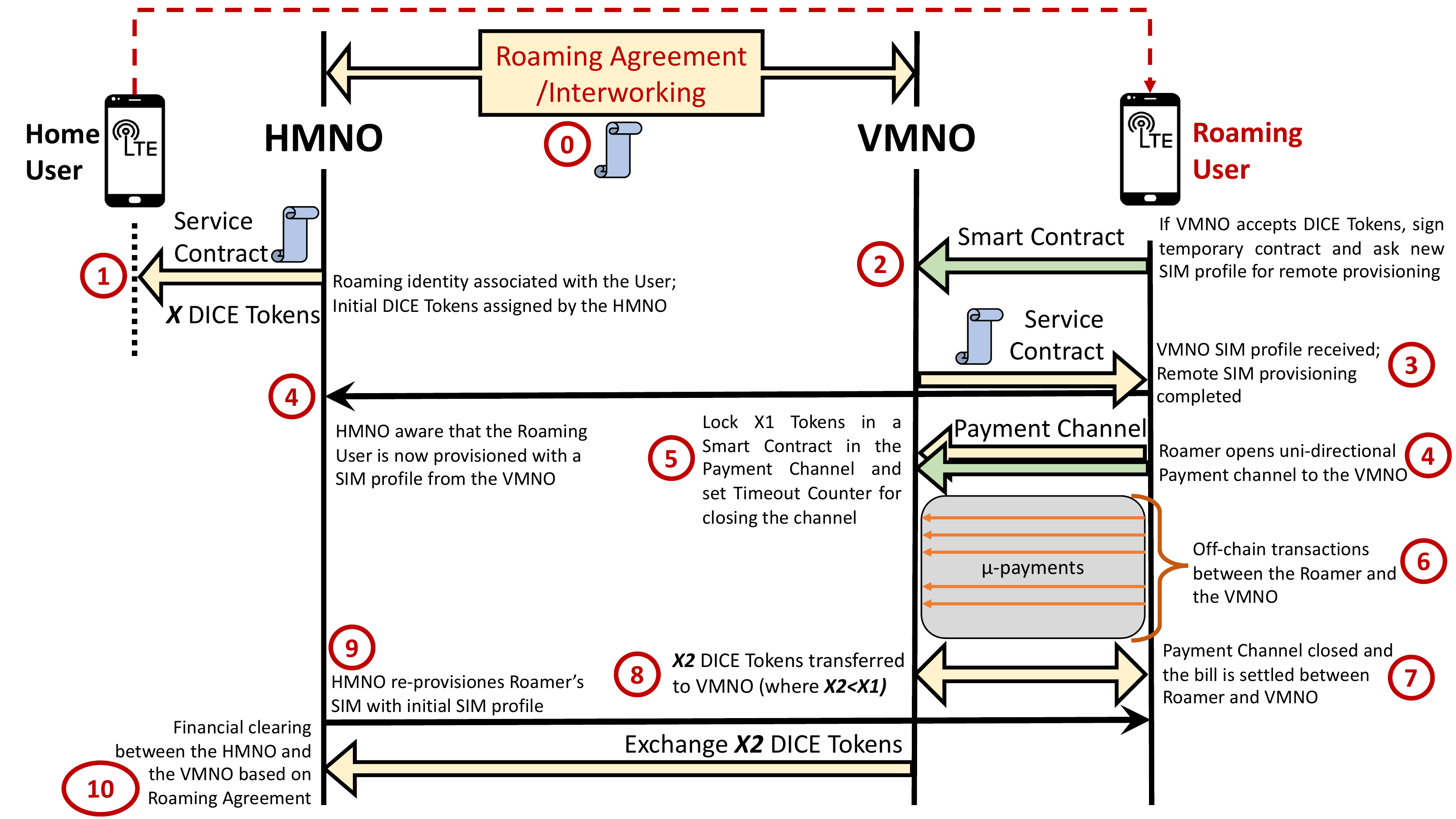}
\vspace{-5mm}
	\caption{A user roaming across MNOs with DICE.}
	\label{fig:dice_mechanism}
	\vspace{-7mm}
\end{figure}

\subsection{DICE Mechanism}
\label{sec:toy_mechanism}
%We provide next a toy example to demonstrate how Alice can use the DICE solution to roam internationally. 
Figure~\ref{fig:dice_mechanism} shows the different steps involved in the DICE protocol and how Alice can use it to obtain service locally in the Visited Country. 
The protocol is the core component of DICE and we define it to resemble the current interaction between roaming partners and roaming users. 

We assume a number of MNOs agree to form the DICE consortium. 
%As a requirement to join the DICE framework, 
Every MNO establishes roaming interworking and agreements with some of the other MNOs in the consortium (Step 0 in Figure~\ref{fig:dice_mechanism}).  
Each of the MNOs in the consortium owns DICE crypto-currency (unique for each MNO) that they can assign to their customers for roaming services. 

When purchasing connectivity from the HMNO (as part of the initial service contract), Alice has the option to opt-in for the DICE schema and use the tokens from the HMNO to buy service when roaming internationally. 
Alice's decision to use DICE generates an initial assignment from the HMNO to Alice's e-wallet of \textit{X} DICE token that Alice can spend while roaming (Step 1 in Figure~\ref{fig:dice_mechanism}). 
This will be a transaction appended to the DICE ledger, which hereinafter remains immutable and tamper-free. 

When traveling to a visited country, Alice attempts to obtain connectivity service from a local MNO that has a previous roaming agreement with her HMNO. 
Through the use of a smart contract that executes on the DICE blockchain the following two conditions are verified: 
(i) Alice verifies whether the selected VMNO accepts the HMNO DICE tokens (Step 2 in Figure~\ref{fig:dice_mechanism}); 
(ii) the VMNO authenticates Alice as a valid user of the HMNO by verifying that her e-wallet contains DICE tokens from the HMNO and by checking the on-chain associated transaction to the initial assignment from the HMNO. 
If these conditions are met, Alice signs a service contract with the VMNO and requests a new profile to remotely provision her \ac{SIM} card (Step 3 in Figure~\ref{fig:dice_mechanism}).
We argue that the \textit{iSIM} technology (integrated \ac{SIM}) enables a key feature of the DICE solution: roaming like at the destination configuration to resemble the LBO roaming.
Remote SIM provisioning allows Alice to remotely change the SIM profile on a deployed SIM without physically changing the SIM itself (Step 4 in Fig.~\ref{fig:dice_mechanism}).  

Once the new SIM profile is configured, Alice opens a uni-directional roaming payment channel towards the VMNO (Step 5 in Figure~\ref{fig:dice_mechanism}).
In order to ensure that Alice pays her bill for the roaming service, tokens have to be locked up as security in a smart contract for the lifetime of the payment channel (Step 6 in Figure~\ref{fig:dice_mechanism}).
Payment channels allow for practically unlimited transfers between the two participants, as long as the net sum of their transfers does not exceed the deposited tokens. 
These transfers can be performed instantaneously and without any involvement of the actual DICE blockchain itself, except for an initial one-time on-chain creation and an eventual closing of the channel. 
Alice uses the payment channel to issue micro-payments to the VMNO periodically for the service the latter offers (Step 7 in Figure~\ref{fig:dice_mechanism}). 
The MVNO associates a time-out counter after each last micro-payment issued by Alice in order to make sure that the payment channel closes when it has been inactive for a given amount of time (e.g., 24 hours). The channel can also be closed from Alice when she configures another SIM profile of an MVNO or the one of its HMNO.
Once the payment channel is closed and the on-chain transaction completes (Step 8 in Figure~\ref{fig:dice_mechanism}), the VMNO receives the DICE tokens from Alice (Step 9 in Figure~\ref{fig:dice_mechanism}). 
In the same time, the HMNO re-provisiones the Alice's SIM with the initial profile (Step 10 in Figure~\ref{fig:dice_mechanism}). 
Finally, the VMNO can proceed to exchange the tokens it has received from Alice for money from the HMNO (Step 11 in Figure~\ref{fig:dice_mechanism}).

The DICE protocol provides near real time billing as the tokens are locked as Alice (the roamer) starts consuming data, and transferred as the channel is closed. 
It also guarantees that there will not be dispute or monetary losses, as Alice can use the service only in exchange of tokens transferred to the visited network.
We note that the procedures associated with the DICE protocol are automated and do not require any human intervention, besides the initial opt-in for the DICE schema and agreement stipulation across MNOs.

\section{System Requirements}
\label{sec:requirements}
% !TeX root = paper.tex
% ================================================================

In this section, we quantify the roaming interactions of an operational MNO in Europe to derive requirements for the DICE system for deployment (Sec.~\ref{sec:deployment}). 
As DICE is based on \ac{DLT}, its feasibility to support the magnitude of the cellular ecosystem is related to the number of transactions per seconds the system can sustain. 
This amount is dominated by the attach and token transfer operations between roamers and \acp{VMNO} (i.e., steps 2 to 8 of Figure~\ref{fig:dice_mechanism}). These account for three transactions to the DICE blockchain, one for the initial smart contract execution (step 2-3), one for opening the payment channel (step 5) and one for closing it (step 8). 
%Transfers over the payment channel do not generate chain transaction as they happen off-chain.
%Other metrics, as the amount of DICE coins to be exchanged between different operators, as well as the number of transactions between users and home network are relevant metrics but do not constitute the main bottleneck. 

\subsection{MNO dataset}
We quantify the system requirements by focusing on the inbound roamers population dynamics of the European MNNO. %of a large operational cellular network in Europe.  
We analyze the population of the MNO over four weeks, from April 4th to April 30th 2019 (consistent with a billing cycle). 
To preserve confidentiality, we present results normalized, anonymized or as order of magnitude. 
As this is a medium-large size MNO (i.e., 30 millions customers), we further approximate the overall DICE system requirements to cater to as much as 800 MNOs worldwide~\cite{gsma_mno}. 

\begin{figure*}[!ht]
	\centering
		\begin{subfigure}{0.33\textwidth}
		\includegraphics[width=\linewidth]{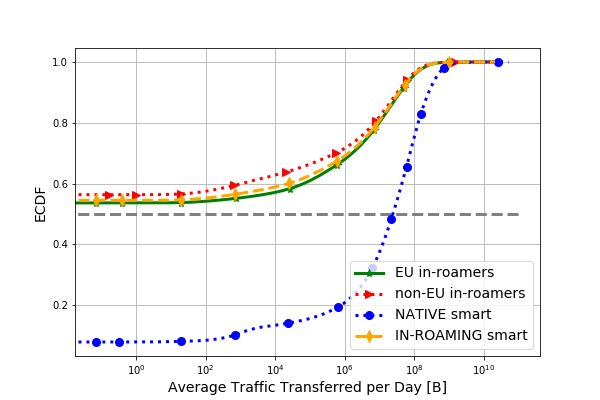}
		\caption{Data traffic per day for inbound roamers.}
		\label{fig:aggregated_bytes}%
	\end{subfigure}%
	\begin{subfigure}{0.33\textwidth}
	\includegraphics[width=.98\linewidth]{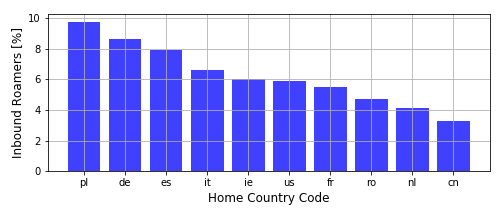}
		\caption{Distribution of Fraction of Inbound Roamers per Home Country.} 
	\label{fig:inroam_per_homeCC}
	\end{subfigure}%
	\begin{subfigure}{0.33\textwidth}
	\includegraphics[width=.98\linewidth]{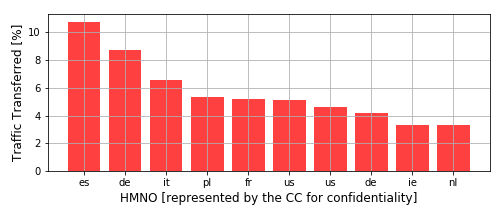}
		\caption{Distribution of Inbound Roamers traffic per Home MNO.} 
	\label{fig:inroam_per_homeMNO}
	\end{subfigure}%
		%\begin{subfigure}{0.33\textwidth}
		%\includegraphics[width=\linewidth]{fig/CDF_TrafficConsumption.png}
		%\caption{CDF of data traffic from inbound roamers grouped per home country.}
		%\label{fig:aggregated_per_homeCC}%
	%\end{subfigure}%
	\vspace{-2mm}
	\caption{Analysis for inbound roaming smartphones for an operational MNO in Europe.}
	\vspace{-5mm}
	\label{fig:inroam_data_consumed}
\end{figure*}

Although the dataset includes smartphones, feature phones, and IoT devices, we use a commercial database provided by GSMA to separate for our analysis the smartphones that people use as their primary devices. 
This latter population subgroup integrates users with different roaming status, including MNO's \emph{native} users (active both in the home country and abroad as outbound roamers) or foreign users that belong to other MNOs (national or international), but who use the radio network of the MNO under analysis (i.e., inbound roamers).
We separate and analyze the inbound roamers in the MNO's network (checking the radio network logs), the outbound roamers (checking the core network 
logs) and the native users of the \ac{MNO} in the home country.  
%To track the radio activity of the inbound roamers, we process logs reporting on activities on IuCS, IuPS, A, and Gb radio interfaces.
%Those carry events generated by the devices connecting to the radio sectors and requesting resources for data communications.
Further we use \acp{xDR} to provide aggregate service usage for data communications. 
%Each record reports the anonymized user ID, MCC and MNC codes for both device SIM and visited country,  timestamp, duration, and bytes consumed. 

%We combine the three data sources to create a daily list of active devices and associated properties and traffic characteristics. 
%Each record in this users catalog reports a device ID, total number of events, calls, bytes seen, \ac{SIM} MCC/MNC, list of
%visited MCC-MNC, list of \ac{APN} strings, device manufacturer, device model, device \ac{OS}.

\subsection{Analysis}

\noindent{\bf Number of roamers. }
We first note that the inbound roamers are the third largest group in the MNO population (17\%), after native users (50\%) and users operating with a SIM card from an MVNO enabled by the incumbent MNO (33\%). 
This population integrates all device types, including \ac{IoT} devices and smartphones. 
Using the information from the MNO dataset, we are able to separate for further analysis only the smartphones (e.g., by verifying if the GSMA database associated an operating system to the devices such as iOS, Android, Windowsphone, Blackberry). 
The total population of inbound roaming \textit{smartphones} per day accounts for 2.5\% of the total number of devices connected to the MNO's radio network on average across the one month period. This percentage represents \emph{hundreds of thousands devices that need to be sustained by DICE for a given visited MNO}.
In the rest of this section, we focus on this subgroup of devices to understand how many transactions they would generate.

\noindent{\bf Number of transactions for newly connected and leaving roamers (step 2-8).}
The number of newly connected inbound roaming smartphones (i.e., who start a new roaming session with the visited MNO during the period of analysis) per day fluctuates with between 10\% and 30\% of the average daily number of inbound roamers connected to the MNO during April 2019.  
%75,000 and 175,000 users divided by 650,000 average daily inbound roamers
Further, the number of leaving inbound roamers (i.e., who finalize their roaming session with the visited MNO during the period of analysis) similarly fluctuates between 10\% and 30\% of the average number of inbound roamers.  We also measured that, in median, the number of days an inbound roamer is active in the visited MNO is approximately 2.5 days.
These results highlight that \emph{tens of thousand of roamers per day that would generate a transaction to execute the initial smart-contract, one to connect to and one to leave a visited network for a given MNO}. Assuming 800 MNOs, this results in few millions operations per day; even assuming these transactions are concentrated in few hours, there would be few thousands operations per second on the blockchain that can be sustained by existing technologies (cf. Sec. ~\ref{sec:deployment}).
% 10,000 to 15,000 smartphones.    

\noindent{\bf Number of transactions for data usages.}
%We next investigate the service usage activity of the inbound roamers in terms of generating billable data communications.
%As described in Sec~\ref{sec:toy_example} each roamer could generate transactions in DICE proportionally to the data consumed. 
We show in Figure~\ref{fig:inroam_data_consumed} the distributions of amount of traffic transferred per day (in Bytes) by the inbound roamers. 
We find that more than 50\% of all inbound roamers are silent roamers, that do not generate data communications. 
We note that the median amount of traffic per day transferred by a non-silent inbound roamer is approximately 1MB 
("IN-ROAMING smart" curve in Figure~\ref{fig:aggregated_bytes}). This translates into an average sum of traffic transferred by one inbound roamer of approximately 2.5MB during a roaming visit.  
This is 10 times smaller the the median amount of traffic transfered by a native user of the MNO ("NATIVE smart" curve in Figure~\ref{fig:aggregated_bytes}). 
We compare the amount of traffic aggregated for inbound roamers from countries in the EU ("EU in-roamers" curve in Figure~\ref{fig:aggregated_bytes}) against countries from outside the EU ("non-EU in-roamers" curve in Figure~\ref{fig:aggregated_bytes}) and find that there is slight increase in the traffic for the EU inbound roamers. Overall, we observe around 10 TB of data generated by roamers. Assuming 100KB as billing granularity as in today's roaming configuration, \emph{there would be hundreds of millions transactions per day for a given visited MNO}. With 800 MNOs, this results in few billions operations per day: this high number of transactions justifies the design choice of performing them off-chain so that they do not constitute a bottleneck (cf. Sec.~\ref{sec:deployment}).

%\dpnote{I would cut from here...}

%For 10\% of all inbound roamers, the duration of their roaming session was higher than 10 days in April 2019. 
%We further aggregated the inbound roamers per home country and show in Figure~\ref{fig:aggregated_per_homeCC} the CDFs of data traffic per day for roamers incoming from Poland, Germany, Spain, US and China. 
%The results highlight the difference in the service usage of different inbound roamers depending on their home country, bringing to attention the potential need of different billing models for the different roaming partners in question. 
%Surprisingly, we note that the ratio of silent roamers from Poland is very high (70\%), more so than the one for inbound roamers from outside the EU (e.g., US or China).  \dpnote{... to here}

\noindent{\bf Number of partner MNOs and countries.}
Over the entire month of April, we find that the MNO provided service to more than 18 million unique inbound roamers, associated to 
approximately 400 MNOs mapping to 188 home countries overall. 
We show in Figure~\ref{fig:inroam_per_homeCC} the distribution of inbound roamers per home country for the top 10 countries as percentages of the total inbound roamers observed over the one-month period. 
The top home countries include Poland, Germany, Spain, Italy and Ireland and account for about 60\% of inbound roamers. 
Although this proves increased mobility within the EU (intuitive as a consequence of "Roam like at Home" EU Regulation~\cite{ec-roaming-charges}), we also observe a large number from non-EU states, such as US or India.  We then report in Figure~\ref{fig:inroam_per_homeMNO} the distribution of inbound roamers traffic per home MNO for the top 10 MNOs. We again observe a long tail distribution with the top 10 MNOs accounting for about 50\% of traffic.
These results highlight that~\emph{few MNOs from few countries will cover the majority of tokens to be exchanged; however, transactions with the long tail of MNOs are required to exchange the remaining tokens}. These metrics do not constitute a bottleneck as token exchanges happen on a long timescale (e.g., weeks or months), but they are relevant to take into account in DICE deployment consideration.

\section{DICE Implementation}
\label{sec:deployment}
% !TeX root = paper.tex
% ================================================================

In this section, we discuss the practical implementation of DICE. 
We examine the deployment requirements that ensure a smooth transition from the current ecosystem to one where DICE is a valid solution. 
Then, we extend the mechanism we presented in Sec.~\ref{sec:toy_mechanism} on several practical aspects for the DICE implementation, feasibility, and charging principles. 
%\dpnote{Add few words saying we look into feasibility}
%we presented a simplified version of the DICE protocol to provide the intuition behind the solution we propose to build. 
%We expand here on several of the procedures we integrate in the DICE protocol.
%We also discuss different design alternatives that insure a smooth transition from the current ecosystem to one where DICE is a valid solution.  
%Finally, we revisit the dimensional requirements we derived in Section~\ref{sec:requirements} and expand here on several practical aspects for the DICE prototype implementation. 

\subsection{Deployment Considerations}

\noindent{\bf DICE Backwards Compatibility with HR Roaming.} 
The protocol in Figure~\ref{sec:toy_mechanism} includes the use of remote SIM provisioning through iSIM to enable LBO roaming.
Though this is the desirable configuration of the DICE protocol, we also ensure compatibility with the current HR roaming status. 
In this latter case, we would skip Step 3 in the DICE protocol diagram in Figure~\ref{fig:dice_mechanism}, and simply rely on the existing roaming configuration implemented by the HMNO. 
Assuming HR roaming, the HMNO is still the entity that controls the roamer's exit point to the Internet and there is no need for a legal contract with the VMNO in the visited country. 
Thus, the ledger accounts in near-real time the activity of the roamer in the VMNO by translating the received service (i.e., transferred traffic) into DICE tokens the roamers pays to the VMNO. This process is compatible with the logic of steps 4-8 in Figure~\ref{fig:dice_mechanism}.

\noindent{\bf Blockchain Solution Feasibility.} 
%\dpnote{I would call blockchain solution and feasibility analysis.}
%The purpose of the DICE blockchain is to allow MNOs to engage in a roaming partnership and to exchange value easily, without the need of a third party to act as a trusted intermediary with the role to verify the interaction between the roaming partners. 
%Blockchain brings the opportunity of using technology to lower uncertainty about the identity of the different entities involved in the roaming transaction (i.e, the roaming user, the home network and the visited network offering roaming services to the roaming user). 
%The notion of uncertainty to which we refer here integrates three main components: identity (not knowing with whom we are dealing), visibility of transactions between any entities and recourse in case something goes wrong. 
%We argue that the blockchain technology can address all of these and enable the exchange of value without trust between the entities. 
%Moreover, smart contracts (or blockchain contracts) are computer programs that act as agreements and which can self-execute on the blockchain. 
%In DICE, we leverage the blockchain contracts to enable the MNO partners to offer service to roaming users and coordinate this exchange. 
%Binding contracts guarantee that the contracts execute without a third party.
%A smart contract in DICE can vary from an MNO offering aggregated service to a user to MNOs registering roaming agreements or a roaming user getting service locally from a visited network. 
%
%\noindent{\bf Consortium Blockchain among MNO.}
We propose the implementation of a consortium blockchain, which has many of the same advantages of a private blockchain, but operates under the leadership of a group (instead of a single entity). 
As opposed to the public blockchain model, which allows any person with an internet connection to participate in the verification of transactions process, a few selected nodes are predetermined. 
%Members of this consortium are all \acp{MNO} using DICE. 
A consortium blockchain solution allows the creation of crypto-currencies with meaning only within the private group (different from public crypto-currencies such as Bitcoin).
It also keeps the control of the blockchain among its members and avoids external entities to influence or attack the system. 
Also, consortium blockchains allow for competing business entities to interact on the same blockchain~\cite{guo2016blockchain} (see privacy and confidentiality considerations).  
%We evaluate here the need of implementing DICE using a specifically tailored blockchain or the potential of using existing blockchain (e.g., Ethereum).

Given the requirements in terms of the number of blockchain transactions per second (TPS) we approximated in Section~\ref{sec:requirements}, we propose building DICE with Hyperledger Fabric~\cite{Androulaki:2018:HFD:3190508.3190538} -- a permissioned blockchain -- as the distributed tamper-proof ledger. 
Hyperledger can accommodate 20,000TPS, while other blockchains such as Ethereum can accommodate only 10-25 TPS or Ripple 1,500 TPS. 
This is enough to process the number of funding transactions required for the creation of the channels for the visiting nodes.

Besides dimensional requirements, Hyperledger is also a suitable technology for the following reasons: 
i) it offers a modular framework that allows for assets to be transferred between blockchain participants (in our case, DICE tokens that account for mobile connectivity service);
ii) it allows the members of the permissioned network to define the asset type, its value and the consensus protocol; 
iii) Hyperledger provides membership identity service that manages participants' identities and authenticates them; 
%It can also enable the use of additional access control lists to provide additional layers of permission;
iv) the ledger consists of a database storing the current state of the network and a log of transactions for tracking each asset's provenience. 

\noindent{\bf Payment channels with off-chain transactions.}
We propose to use unidirectional payment channels that allow the transfer of coins from the roamer to the VMNO, based on the traffic exchanged by the visiting node.  
In particular, we propose a granularity of 100KB, meaning that the roaming user will transfer coins for every 100KB of traffic transferred through the VMNO. 
This is the typical billing granularity currently used for roaming.
To enable scalability of this charging model while meeting our requirements, DICE relies on payment channels based simple on hashed time-locks contracts that enable off-chain transactions between the roaming user and the VMNO~\cite{raiden}.
%Specifically, we use payment channels based on hashed time-locks contracts that are enough for these purposes \dpnote{enough in which sense?} and provide the scalability required. 
It follows that each visiting node will require the creation of one on-chain funding transaction for creating the channel and all the transactions paying for the traffic exchanged will happen off-line.

\noindent{\bf Roaming Privacy and Confidentiality.}
The usage of a distributed ledger implies that every party involved in the network would have access to all transactions, even the ones where they were not involved in the transactions themselves. 
This brings a challenge for telco businesses that are competing in the same space and do not want to reveal company secrets to their competitors. 
This also impacts privacy of end users that could be tracked across MNOs.
Hyperledger Fabric solves the former problem by offering private channels, which are restricted messaging paths that offer privacy for specific subsets. 
In this case, all data is invisible to members not granted specific access. 
To further obfuscate the number of customers an MNO has active in DICE and preserve user privacy, we envision that the \ac{HMNO} can assign to the roamer multiple identities (or e-wallets), to which the latter attaches different amounts of DICE tokens. 

\subsection{DICE Charging Principles}

In the last step of the protocol depicted in Fig.~\ref{fig:dice_mechanism}, the VMNO triggers the financial clearing procedure by exchanging the DICE tokens from its inbound roamers with the corresponding HMNOs. 
To avoid potential speculations on exchanging these token to money, we implement in DICE a prior check for the provenience of the tokens being exchanged. 
Specifically, for an HMNO to accept the tokens from a partner VMNO, the latter must prove that it received the tokens from a customer of the HMNO in exchange for its service. 
In other words, the HMNO will convert to fiat money the DICE coins that went from it own customers to the VMNO. 

This final financial settlement step also implies the use of a specific billing model, one that the roaming partners likely agree upon in the Roaming Agreement they sign (Step 0 in Figure~\ref{fig:dice_mechanism}).
This can accommodate different charging models.
In the past decades the wholesale roaming charge models for data usage have remained in the retail-like charge models, which are applied to each visitor who roamed in the VMNO (per session per IMSI).
With increasing roaming data usage there is the trend to use a simplified IOT (Inter Operator Tariff) model such as the “Fixed” or “Per-Unit” charge model which is applied to the total volume of data usage with or without a post-discount adjustment.
We argue that DICE offers maximum flexibility of implementing dynamics charging models at different granularity levels.

\section{Discussion and Future Work}
\label{sec:discussion}
% !TeX root = paper.tex
% ================================================================

In \ac{DLT} we may be witnessing a potential explosion of creative potential that catalyzes exceptional levels of innovation.  
The technology has the capacity to deliver a new kind of trust to a wide range of services, including the largely opaque cellular roaming ecosystem.
We argue that DICE will enable a much-needed communication loop between the mobile connectivity provider and the subscribers, allowing the latter to gain control over their connectivity and request certain levels of service. 
Also, \acp{MNO} can form alliances to share their infrastructure and jointly fulfill the requirements of subscribers. 
The DICE open and competitive framework provides the foundations for growth, innovation and affordable access for the end users.

 In continuing the work we outlined in this paper, we aim for a proof-of-concept implementation of DICE. 
 In this framework, we then plan to explore multiple open questions related to the performance of the DICE solution, privacy of the end-user and security of the system. 
 We envision that we can extend DICE to integrate different other functionality, such as real-time fraud detection and anomaly detection.

%\balance
\bibliographystyle{abbrv} 
\begin{small}
\bibliography{references}
\end{small}

\end{document}